\documentclass[english,aps,floats,twocolumn,showpacs,prl,
nofootinbib]{revtex4}
\usepackage{pslatex}
\usepackage[T1]{fontenc}
\usepackage[latin1]{inputenc}
\usepackage{graphicx}
\usepackage{epsfig}
{
{\newcommand{\gsim}{\mbox{\raisebox{-.6ex}{~$\stackrel{>}{\sim}$~}}} 
\newcommand{\be}{\begin{equation}}
\newcommand{\ee}{\end{equation}}
\newcommand{\bea}{\begin{eqnarray}}
\newcommand{\eea}{\end{eqnarray}}

\begin{document}
\title{Predictive model for dark matter, dark energy, neutrino masses 
and leptogenesis at the TeV scale}
\author{Narendra Sahu}
\author{Utpal Sarkar}
\affiliation{Theory Division, Physical Research Laboratory,
Navarangpura, Ahmedabad, 380 009, India}
\begin{abstract}
We propose a new mechanism of TeV scale leptogenesis where the 
chemical potential of right-handed electron is passed on to 
the $B-L$ asymmetry of the Universe in the presence of sphalerons. 
The model has the virtue that the origin of neutrino masses are 
independent of the scale of leptogenesis. As a result, the model 
could be extended to explain {\it dark matter, dark energy, neutrino 
masses and leptogenesis at the TeV scale}. The most attractive feature 
of this model is that it predicts a few hundred GeV triplet Higgs 
scalar that can be tested at LHC or ILC.    

\end{abstract}
\maketitle
\section{Introduction} 
In the canonical seesaw models~\cite{canonical_seesaw} 
the physical neutrino masses are largely suppressed by the scale 
of lepton (L) number violation, which is also the scale of leptogenesis. 
The observed baryon (B) asymmetry and the low energy neutrino oscillation 
data then give a lower bound on the scale of leptogenesis to be $\sim 10^9$ 
GeV~\cite{di_bound}. Alternately in the triplet seesaw models~\cite{
tripletseesaw} it is equally difficult to generate $L$-asymmetry at 
the TeV scale because the interaction of $SU(2)_L$ triplets with the 
gauge bosons keep them in equilibrium up to a very high scale $\sim 10^{10}$ 
GeV~\cite{ma&sarkar_prl}. However, in models of extra dimensions~\cite{
ma_raidal_sarkar} and models of dark energy~\cite{ma&sarkar_plb} the masses 
of the triplet Higgs scalars could be low enough for them to be accessible 
in LHC or ILC, but in those models leptogenesis is difficult. Even in the 
left-right symmetric models in which there are both right-handed neutrinos 
and triplet Higgs scalars contributing to the neutrino masses, it is
difficult to have triplet Higgs scalars in the range of LHC or
ILC~\cite{sahu&sarkar_prd}. It may be possible to have resonant 
leptogenesis~\cite{singlet_resonant} with light triplet Higgs 
scalars~\cite{triplet_resonant}, but the resonant condition requires 
very high degree of fine tuning.

In this paper we introduce a new mechanism of leptogenesis at the 
TeV scale. We ensure that the lepton number violation required for 
the neutrino masses does not conflict with the lepton number violation 
required for leptogenesis. This led us to propose a model which is 
capable of explaining dark matter, dark energy, neutrino masses 
and leptogenesis at the TeV scale. Moreover, the model predicts a 
few hundred GeV triplet Higgs whose decay through the same sign 
dilepton signal could be tested either through the $e^\pm e^\mp$ 
collision at linear collider or through the $pp$ collision at LHC.

\section{The model} 
In addition to the quarks, leptons and the usual 
Higgs doublet $\phi \equiv (1,2,1)$, we introduce two triplet Higgs 
scalars $\xi \equiv (1,3,2)$ and $\Delta \equiv (1,3,2)$, two 
singlet scalars $\eta^- \equiv (1,1,-2)$ and $T^0 \equiv (1,1,0)$, 
and a doublet Higgs $\chi\equiv (1,2,1)$. The transformations of 
the fields are given under the standard model (SM) gauge group 
$SU(3)_c \times SU(2)_L \times U(1)_Y$. There are also three heavy 
singlet fermions $S_a \equiv (1,1,0), a = 1,2,3$. A global symmetry 
$U(1)_X$ allows us to distinguish between the $L$-number violation 
for neutrino masses and the $L$-number violation for leptogenesis. 
Under $U(1)_X$ the fields $\ell_{iL}^T \equiv (\nu, e)_{iL} \equiv 
(1,2,-1)$, $e_{iR} \equiv (1,1,-2)$, $\eta^-$ and $T^0$ carry a quantum 
number 1, $\Delta$, $S_a$, $a=1,2,3$ and $\phi$ carry a quantum 
number zero while $\xi$ and $\chi$ carry quantum numbers -2 and 2 
respectively. We assume that $M_\xi \ll M_\Delta$ while both $\xi$ 
and $\Delta$ contribute equally to the effective neutrino masses. 
Moreover, if neutrino mass varies on the cosmological time scale 
then it behaves as a negative pressure fluid and hence explains 
the accelerating expansion of the present Universe~\cite{dark_energy}
\footnote{Connection between neutrino mass and dark energy, which 
is required for accelerating expansion of the Universe, in large 
extradimension scenario is discussed in ref.~\cite{cliffetal}}. With 
a survival $Z_2$ symmetry, the neutral component of $\chi$ represents 
the candidate of dark matter~\cite{darkmatter}. 

Taking into account of the above defined quantum numbers we now 
write down the Lagrangian symmetric under $U(1)_X$. The terms in the
Lagrangian, relevant to the rest of our discussions, are given by
\bea
-{\cal L } &\supseteq & f_{ij} \xi \ell_{iL} \ell_{jL} + \mu (A) \Delta^\dagger
\phi \phi + M_\xi^2  \xi^\dagger \xi + M_\Delta^2 \Delta^\dagger
\Delta \nonumber\\
&& + h_{ia} \bar e_{iR} S_a \eta^- + M_{sab} S_a S_b + y_{ij}\phi
\bar \ell_{iL} e_{jR}+ M_T^2 T^\dagger T \nonumber\\
&&+ \lambda_T |T|^4 + \lambda_\phi |T|^2 |\phi|^2 + \lambda_\chi |T|^2
|\chi|^2 + f_T \xi \Delta^\dagger T T\nonumber\\
&& + \lambda_{\eta \phi}|\eta^-|^2 |\phi|^2 + \lambda_{\eta \chi}
|\eta^-|^2|\chi|^2 + V_{\phi \chi} + h.c.\,,
\label{lagrangian}
\eea
where $V_{\phi \chi}$ constitutes all possible quadratic and quartic 
terms symmetric under $U(1)_X$. The typical dimension full coupling 
$\mu(A)=\lambda A$, $A$ being the acceleron field\footnote{The origin of
this acceleron field is beyond the scope of this paper. See for example 
ref.~\cite{pascosetal.06}.}, which is responsible for the accelerating 
expansion of the Universe. We introduce the $U(1)_X$ symmetry breaking 
soft terms
\begin{equation}
-\mathcal{L}_{soft} = m_T^2 T T + m_\eta \eta^- \phi \chi + h.c. \,.
\label{soft-terms}
\end{equation} 
If $T$ carries the $L$-number by one unit then the first term 
explicitly breaks $L$-number in the scalar sector. The second term 
on the other hand conserves $L$-number if $\eta^-$ and $\chi$ possess 
equal and opposite $L$-number\footnote{If $\eta^-$ does not possess 
any $L$-number then the interaction of $S_a$ explicitly breaks 
$L$-number and hence the decay of lightest $S_a$ gives rise to a 
net $L$-asymmetry as in the case of right handed neutrino 
decay~\cite{fukugita.86}.}. This leads to the interactions of the fields 
$S_a, i=1,2,3$ to be $L$-number conserving. As we shall discuss later, 
this can generate the $L$-asymmetry of the universe, while the neutrino 
masses come from the $L$-number conserving interaction term 
$\Delta^\dagger \xi T T $ after the field $T$ acquires a $vev$.

\section{Neutrino masses}
The Higgs field $\Delta$ acquires a very small 
vacuum expectation value ($vev$) 
\begin{equation}
\langle \Delta \rangle = -\mu (A) {v^2 \over M_\Delta^2}\,,
\end{equation}
where $v = \langle  \phi \rangle$, $\phi$ being the SM Higgs doublet. 
However, we note that the field $\xi$ does not acquire a $vev$ at the 
tree level. 

The scalar field $T$ acquires vev at a few TeV, which then induces a 
small $vev$ to the scalar field $\xi$. The Goldstone boson corresponding 
to the $L$-number violation, the would be Majoron, and the Goldstone 
boson corresponding to $U(1)_X$ symmetry will have a mass of the order 
of a few TeV and will not contribute to the $Z$ decay width. The $vev$ 
of the field $\xi$ would give a small Majorana mass to the neutrinos.

The $vev$ of the singlet field $T$ gives rise to a mixing between 
$\Delta$ and $\xi$ through the effective mass term
\begin{equation}
-{\cal L }_{\Delta \xi} = m_{s}^2 \Delta^\dagger \xi ,
\end{equation}
where the mass parameter $m_s = \sqrt{f_T \langle T \rangle^2}$ is of 
the order of TeV, similar to the mass scale of $T$. The effective 
couplings of the different triplet Higgs scalars, which give the 
$L$-number violating interactions in the left-handed sector, are 
then given by
\begin{eqnarray}
-{\cal L }_{\nu - mass} &=& f_{ij} \xi \ell_i \ell_j + \mu (A) {m_s^2 \over 
M_\Delta^2} \xi^\dagger \phi \phi + f_{ij} {m_s^2 \over M_\xi^2} 
\Delta \ell_i \ell_j \nonumber\\
&& + \mu (A) \Delta^\dagger \phi \phi + h.c.\,.
\label{flavour_vio}
\end{eqnarray}
The field $\xi$ then acquires an induced $vev$,
\be
\langle \xi \rangle = -\mu (A) {v^2 m_s^2 \over M_\xi^2 M_\Delta^2}\,.
\label{xi_vev}
\ee
The $vev$s of both the fields $\xi$ and $\Delta$ will contribute to
neutrino mass by equal amount and thus the neutrino mass is given 
by
\begin{equation}
{m}_{\nu} = - f_{ij} \mu (A) {v^2 m_s^2 \over M_\xi^2 M_\Delta^2}\,.
\label{neutrino_mass}
\end{equation}

Since the absorptive part of the off-diagonal one loop self energy 
terms in the decay of triplets $\Delta$ and $\xi$ is zero, their 
decay can't produce any $L$-asymmetry even though their decay 
violate $L$-number. However, the possibility of erasing any 
pre-existing $L$-asymmetry through the $\Delta L=2$ processes 
mediated by $\Delta$ and $\xi$ should not be avoided unless their 
masses are very large and hence suppressed in comparison to the 
electroweak breaking scale. In particular, the important erasure 
processes are: 
\be
\ell \ell \leftrightarrow \xi \leftrightarrow \phi \phi ~~~~{\mathrm and}~~~~ 
\ell \ell \leftrightarrow \Delta \leftrightarrow \phi \phi\,.
\ee
If $m_s^2\ll M_\Delta^2$ then the $L$-number violating processes mediated 
through $\Delta$ and $\xi$ are suppressed by $(m_s^2/M_\xi^2 M_\Delta^2)$ 
and hence practically don't contribute to the above erasure processes. 
Thus a fresh $L$-asymmetry can be produced at the TeV scale. 

\section{Leptogenesis}
We introduce the following two cases for 
generating $L$-asymmetry which is then transferred to the required 
$B$-asymmetry of the Universe.\\

{\it Case-I}:: The explicit $L$-number violation 

First we consider 
the case where $L$-number is explicitly broken in the singlet sector. 
This is possible if $\eta^-$, and hence $\chi$, does not possess any 
$L$-number. Therefore, the decays of the singlet fermions $S_a$, 
$a=1,2,3$ can generate a net $L$-asymmetry of the universe through
\begin{eqnarray}
S_a &\to & e_{iR}^- + \eta^+ \nonumber \\
& \to & e_{iR}^+ + \eta^- \,.\nonumber
\end{eqnarray}
We work in the basis, in which $M_{sab}$ is diagonal and $M_3 > 
M_2 > M_1$, where $M_a = M_{saa}$. Similar to the usual right-handed 
neutrino decays generating $L$-asymmetry~\cite{fukugita.86}, there 
are now one-loop 
self-energy and vertex-type diagrams that can interfere with the 
tree-level decays to generate a CP-asymmetry. The decay of the field 
$S_1$ can now generate a CP-asymmetry
\begin{eqnarray}
\epsilon &=& - \sum_i \left[ {\Gamma(S_1 \to e_{iR}^- \eta^+)
- \Gamma(S_1 \to e_{iR}^+ \eta^-) \over \Gamma_{tot} (S_1)}
\right] \nonumber \\
&\simeq & {1 \over 8 \pi} {M_1 \over M_2} {{\rm Im} [(h h^\dagger)_{i1} 
(h h^\dagger)_{i1}] \over \sum_a |h_{a1}|^2}\,.
\label{cp_asymmetry}
\end{eqnarray}
Thus an excess of $e_{iR}$ over $e^c_{iR}$ is produced in the thermal 
plasma. This will be converted to an excess of $e_{iL}$ over $e^c_{iL}$ 
through the t-channel scattering process $e_{iR}e^c_{iR}\leftrightarrow 
\phi^0 \leftrightarrow e_{iL}e^c_{iL}$. This can be understood as follows. 
Let us define the chemical potential associated with $e_R$ field as 
$\mu_{eR}=\mu_0 + \mu_{BL}$, where $\mu_{BL}$ is the chemical potential 
contributing to $B-L$ asymmetry and $\mu_0$ is independent of $B-L$. At 
equilibrium thus we have 
\be
\mu_{e_L} = \mu_{e_R}+\mu_\phi = \mu_{BL} + \mu_0+\mu_\phi\,. 
\ee
We see that $\mu_{eL}$ is also associated with the 
same chemical potential $\mu_{BL}$. Hence the $B-L$ asymmetry produced 
in the right-handed sector will be transferred to the left-handed sector. 
A net baryon asymmetry of the universe is then produced through the 
sphaleron transitions which conserve $B-L$ but violate $B+L$. Since the 
source of $L$-number violation for the this asymmetry is different 
from the neutrino masses, there is no bound on the mass scale of $S_1$
from the low energy neutrino oscillation data. Therefore, the mass
scale of $S_1$ can be as low as a few TeV. Note that the mechanism
for $L$-asymmetry proposed here is different from an earlier proposal
of right handed sector leptogenesis~\cite{frigerio_ma}. The survival 
asymmetry in the $\eta$ fields is then transferred to $\chi$ fields 
through the trilinear soft term introduced in Eq. (\ref{soft-terms}).

{\it Case-II}:: Conserved $L$-number 

We now consider the case where 
$L$-number is conserved in the singlet sector. This is possible 
if $\eta^- (\eta^+)$ possesses a $L$-number exactly opposite to that of 
$e_R^+ (e_R^-)$. Therefore, the decays of the singlet fermions $S_a$, 
$a=1,2,3$ can not generate any $L$-asymmetry. However, it produces an 
equal and opposite asymmetry between $\eta^- (\eta^+)$ and $e_R^+ 
(e_R^-)$ fields as given by Eq. (\ref{cp_asymmetry}). If these two 
asymmetries cancel with each other then there is no left behind 
$L$-asymmetry. However, as we see from the Lagrangians (\ref{lagrangian}) 
and (\ref{soft-terms}) that none of the interactions that can 
transfer the $L$-asymmetry from $\eta^-$ to the lepton doublets while 
$e_R$ is transferring the $L$-asymmetry from the singlet sector to the 
usual lepton doublets through $\phi \bar{\ell}_L e_R$ coupling. Note that 
the coupling, through which the asymmetry between $\eta^-$ and 
$e_R^+$ produced, is already gone out of thermal equilibrium. 
So, it will no more allow the two asymmetries to cancel with 
each other. The asymmetry in the $\eta$ fields is finally transferred 
to the $\chi$ fields through the trilinear soft term introduced in 
Eq. (\ref{soft-terms}).   

\section{Dark matter} 
As the universe expands the temperature of the 
thermal bath falls. As a result the heavy fields $\eta^-$ and $T^0$ 
are annihilated to the lighter fields $\phi$ and $\chi$ as they are 
allowed by the Lagrangians (\ref{lagrangian}) and (\ref{soft-terms}). 
Notice that there is a $Z_2$ symmetry of the Lagrangians (\ref{lagrangian}) 
and (\ref{soft-terms}) under which $S_a, a=1,2,3$, $\eta^-$ and $\chi$ are 
odd while all other fields are even. Since the neutral component of 
$\chi$ is the lightest one it can be stable because of $Z_2$ symmetry. 
Therefore, the neutral component of $\chi$ behaves as a dark 
matter. 

After $T$ gets a vev the effective potential describing the 
interactions of $\phi$ and $\chi$ can be given by 
\bea
V (\phi, \chi) &=& \left( -m_\phi^2+ {\lambda_\phi\over f_T}m_s^2 \right) 
|\phi|^2 + \left( m_\chi^2 + {\lambda_\chi \over f_T} m_s^2\right) 
|\chi|^2\nonumber\\ 
&&+\lambda_1|\phi|^4 +\lambda_2 |\chi|^4 +\lambda_3 |\phi|^2 |\chi|^2 
+ \lambda_4 |\phi^\dagger \chi|^2\,,
\label{chi_phi_pot}
\eea
where we have made use of the fact that $m_s = \sqrt{f_T \langle T 
\rangle^2}$ and $\lambda_\phi$, $\lambda_\chi$ are the quartic couplings
of $T$ with $\phi$ and $\chi$ respectively. For $m_\phi^2 > \left( 
{\lambda_\phi\over f_T} \right)m_s^2 >0$ and $m_\chi^2 , \left( \frac{
\lambda_\chi}{f_T} \right) m_S^2  > 0$ the minimum of 
the potential is given by 
\be
\langle \phi \rangle =\pmatrix{
0\cr
v } ~~~~~{\mathrm and}~~~~~  \langle \chi \rangle=\pmatrix{ 
0\cr
0} \,. 
\ee
The $vev$ of $\phi$ gives masses to the SM fermions and gauge bosons. 
The physical mass of the SM Higgs is then given by $m_h = 
\sqrt{4\lambda_1 v^2}$. The physical mass of the real and 
imaginary parts of the neutral component of $\chi$ field are almost same 
and is given by 
\begin{equation}
m_{\chi_{R,I}^0}^2= m_\chi^2+{\lambda_\phi\over f_T}m_s^2+
(\lambda_3+\lambda_4) v^2\,.
\end{equation}
Since $\chi$ is odd under the surviving $Z_2$ symmetry it can't decay 
to any of the conventional SM fields and hence the neutral component 
of $\chi$ constitute the dark matter component of the Universe. Above 
their mass scales $\chi_{R,I}^0$ are in thermal equilibrium through 
the interactions: 
$\lambda_2{\chi_{R,I}^0}^4$ and 
$(\lambda_3+\lambda_4){\chi_{R,I}^0}^2 h^2$. Assuming that 
$m_{\chi_{R,I}^0} < m_W, m_h$ the direct annihilation of a pair 
of $\chi_{R,I}^0$, below their mass scale, to SM Higgs is kinematically 
forbidden. However, a pair of $\chi_{R,I}^0$ can be annihilated 
to the SM fields: $f \bar{f}, W^+ W^-, ZZ, gg, hh\cdots$ through 
the exchange of neutral Higgs $h$. The corresponding scattering 
cross-section in the limit $m_{\chi_{R,I}^0}< m_W, m_h$ is given 
by~\cite{cliff_npb.01}
\begin{equation}
\sigma_{h}|v| \simeq \frac{\lambda^2 m_{\chi_{R,I}^0}^2}{m_h^4}\,,
\label{crossection}
\end{equation} 
where $\lambda=(\lambda_3+\lambda_4)$. 

We assume that at a temperature $T_D$, $\Gamma_{ann}/H(T_D)\simeq 1$, 
where $T_D$ is the temperature of the thermal bath when $\chi_{R,I}^0$ 
got decoupled and 
\be
H(T_D)=1.67 g_*^{1/2}(T_D^2/M_{pl})
\ee
is the corresponding Hubble expansion parameter with $g_*\simeq 100$ being the 
effective number of relativistic degrees of freedom. Using 
Eq. (\ref{crossection}) the rate of annihilation of $\chi_{R,I}^0$ to the 
SM fields can be given by $\Gamma_{ann}= n_{\chi^0}\langle \sigma_h |v|
\rangle$, where $n_{\chi^0}$ is the density of $\chi_{R,I}^0$ at the 
decoupled epoch. Using the fact that $\Gamma_{ann}/H(T_D)\simeq 1$ one 
can get~\cite{sahu&yajnik_plb.06} 
\begin{equation}
z_D\equiv \frac{m_{\chi_{R,I}^0}}{T_D}\simeq \ln \left[ 
\frac{N_{ann} \lambda^2 m_{\chi_{R,I}^0}^3 M_{pl}}{1.67 g_*^{1/2} 
(2\pi)^{3/2}m_h^4} \right]\,,
\end{equation} 
where $N_{ann}$ is the number of annihilation channels which we 
have taken roughly to be 10. Since the $\chi_{R,I}^0$ are stable in the 
cosmological time scale we have to make sure that it should not 
over-close the Universe. For this we calculate the energy 
density of $\chi_{R,I}^0$ at the present epoch. The number density 
of $\chi_{R,I}^0$ at the present epoch 
is given by 
\begin{equation}
n_{\chi_{R,I}^0} (T_0) = \left( T_0/T_D \right)^3 n_{\chi_{R,I}^0} (T_D)\,,
\end{equation} 
where $T_0=2.75^\circ k$, the temperature of present Cosmic Microwave 
Background Radiation. We then calculate the energy density at present 
epoch, 
\begin{eqnarray}
\rho_{\chi_{R,I}^0} &=& \left( \frac{0.98\times 10^{-4} eV} {cm^3}\right)
\nonumber\\ 
&& \frac{1}{N_{ann} \lambda^2} \frac{(m_h/GeV)^4}{ 
(m_{\chi_{R,I}^0}/GeV)^2 }[1+\delta]\,,
\end{eqnarray} 
where $\delta\ll 1$. The critical energy density of the present 
Universe is 
\be 
\rho_c =  3 H_0^2/8 \pi G_N \equiv 10^4 h^2 eV/cm^3\,.
\ee
\begin{figure}
\begin{center}
\epsfig{file=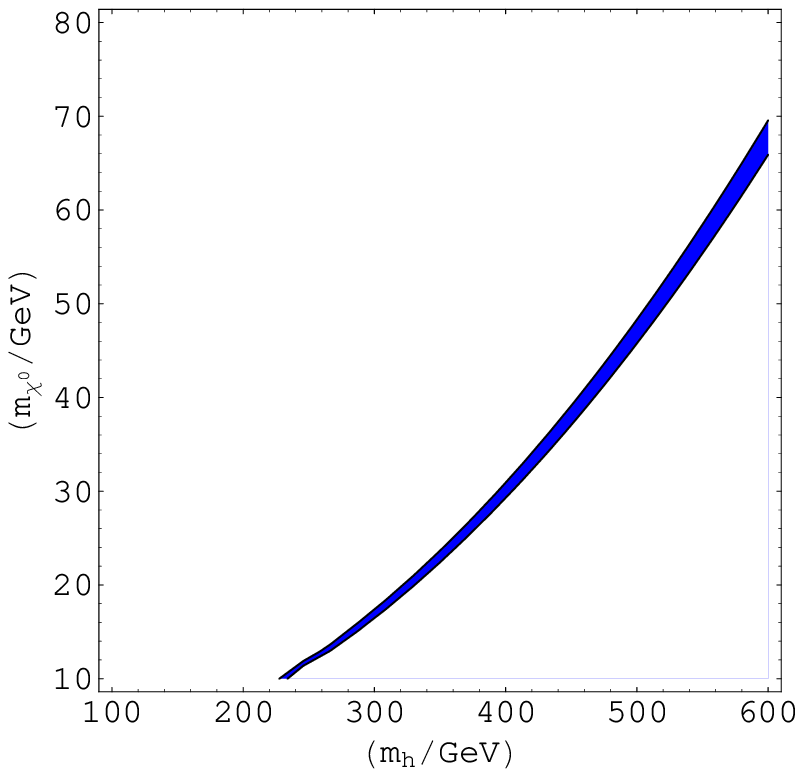, width=0.35\textwidth}
\epsfig{file=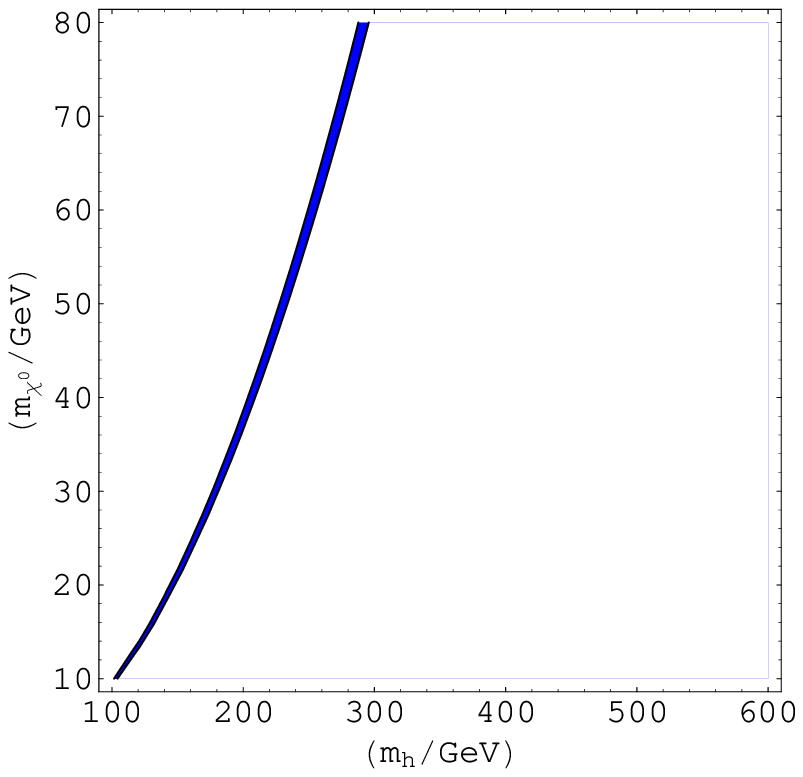, width=0.35\textwidth}
\caption{The allowed region of dark matter at the $1\sigma$ C.L. is shown
in the plane of $m_h$ versus $m_{\chi^{0}}$ with $\lambda^2=0.5$ (upper) 
and $ \lambda^2=0.1$ (bottom).}
\label{figure-2}
\end{center}
\end{figure}
At present the contribution of dark matter to the critical energy density 
of the Universe is precisely given by $\Omega_{DM} h^2=0.111\pm 
006$~\cite{pdg}. Assuming that $\chi_{R,I}^0$ is a candidate of dark 
matter we have shown, in fig. (\ref{figure-2}), the allowed masses 
of $\chi_{R,I}^0$ up to 80 GeV for a wide spectrum of SM Higgs masses. 

\section{Dark energy and neutrino} 
It has been observed that the present 
Universe is expanding in an accelerating rate. This can be attributed 
to the dynamical scalar field $A$~\cite{quintessence}, which evolves 
with the cosmological time scale. If the neutrino mass arises from an 
interaction with the acceleron field, whose effective potential changes 
as a function of the background neutrino density then the observed 
neutrino masses can be linked to the observed acceleration of the 
Universe~\cite{dark_energy}.    

Since the neutrino mass depends on $A$, it varies on the 
cosmological time scale such that the effective neutrino mass is 
given by the Lagrangian 
\begin{equation}
-\mathcal{L}= \left[ f_{ij} \mu (A) {v^2 m_s^2 \over M_\xi^2 M_\Delta^2} 
\nu_i \nu_j+ h.c. \right] + V_0 \,,
\label{varying_mass}
\end{equation}
where $V_0$ is the acceleron potential. A typical form of the 
potential is given by~\cite{ma&sarkar_plb}
\begin{equation}
V_0 = \Lambda^4 \ln \left( 1+ |\bar{\mu}| \mu (A) | \right)\,,
\label{scalar_potential}
\end{equation}
The two terms in the above Lagrangian (\ref{varying_mass}) acts in opposite 
direction such that the effective potential 
\begin{equation}
V (m_\nu)= m_\nu n_\nu + V_0 (m_\nu) 
\end{equation}
today settles at a non-zero positive value. From the above effective 
potential we can calculate the equation of state 
\begin{equation}
w = -1 + \left[ \Omega_\nu/(\Omega_\nu+\Omega_A) \right]\,,
\end{equation}
where $w$ is defined by $V\propto R^{-3(1+w)}$. At present the contribution 
of light neutrinos having masses varying from $5\times 10^{-4}$ eV to 1 MeV 
to the critical energy density of the Universe is $\Omega_\nu\leq 0.0076/h^2$
~\cite{pdg}. Hence one effectively gets $w\simeq -1$. Thus the mass varying 
neutrinos behave as a negative pressure fluid as the dark energy. For 
naturalness we chose $\frac{\mu (A) m_s^2}{M_\Delta^2} \sim 1$ eV such 
that $M_\xi$ can be a few hundred GeV to explain the sub-eV neutrino masses, 
and $\Lambda \sim 10^{-3}$ eV such that the varying neutrino mass can be 
linked to the dark energy component of the Universe.   

\section{Collider signature of doubly charged particles} 
The doubly charged component of the light triplet Higgs $\xi$ can be 
observed through its decay into same sign dileptons~\cite{collider_signature}. 
Since $M_\Delta \gg M_\xi$, the production of $\Delta$ particles in 
comparison to $\xi$ are highly suppressed. Hence it is worth looking
for the signature of $\xi^{\pm\pm}$ either at LHC or ILC. 
From Eq. (\ref{flavour_vio}) one can see that the decay $\xi^{\pm \pm} 
\rightarrow \phi^\pm \phi^\pm$ are suppressed since the decay rate involves 
the factor $\frac{\mu (A) m_s^2}{M_\Delta^2} \sim 1$ eV. While the decay
mode $\xi^{\pm\pm} \rightarrow h^\pm W^\pm $ is phase space suppressed,
the decay mode $\xi^{\pm\pm} \rightarrow W^\pm W^\pm $ is suppressed
because of the vev of $\xi$ is small which is required for sub-eV neutrino
masses as well as to maintain the $\rho$ parameter of SM to be unity. 
Therefore, once it is produced, $\xi$ mostly decay through the same sign 
dileptons: $\xi^{\pm \pm} \rightarrow \ell^\pm \ell^\pm$. Note that the 
doubly charged particles can not couple to quarks and therefore the SM 
background of the process $\xi^{\pm\pm}\rightarrow \ell^\pm \ell^\pm$ is 
quite clean and hence the detection will be unmistakable. From 
Eq. (\ref{flavour_vio}) the decay rate of 
the process $\xi^{\pm\pm}\rightarrow \ell^\pm \ell^\pm$ is given by 
\be
\Gamma_{ii}=\frac{|f_{ii}|^2}{8\pi} M_{\xi^{++}} ~~~~{\mathrm and }~~~~ 
\Gamma_{ij}=\frac{|f_{ij}|^2}{4\pi} M_{\xi^{++}}\,, 
\ee
where $f_{ij}$ are highly 
constrained from the lepton flavor violating decays. From the observed 
neutrino masses we have $f_{ij}x\sim 10^{-12}$ where 
$x=(\langle \xi \rangle/v)$. If $f_{ij}\gsim x$ then from the lepton 
flavor violating decay $\xi^{\pm\pm}\rightarrow \ell^\pm_{i} 
\ell^\pm_{j} $ one can study the pattern of neutrino masses and 
mixing~\cite{chun_plb}.       

\section{Conclusions} 
We introduced a new mechanism of leptogenesis in 
the singlet sector which allowed us to extend the model to explain 
dark matter, dark energy, neutrino masses and leptogenesis at the 
TeV scale. This scenario predicts a few hundred GeV triplet scalar 
which contributes to the neutrino masses. This makes the model 
predictable and it will be possible to verify the model at ILC or LHC 
through the same sign dilepton decay of the doubly charged particles. 
This also opens an window for studying neutrino mass spectrum in 
the future colliders (LHC or ILC). Since the lepton number violation 
required for lepton asymmetry and neutrino masses are different, 
leptogenesis scale can be lowered to as low as a few TeV. 


\begin{thebibliography}{99}

\bibitem{canonical_seesaw} P. Minkowski, Phys. Lett. {\bf B 67}, 421 (1977);
M.~Gell-Mann, P.~Ramond and R.~Slansky in {\it Supergravity} (P.~van 
Niewenhuizen and D.~Freedman, eds), (Amsterdam), North Holland, 1979; 
T.~Yanagida in {\it Workshop on Unified Theory and Baryon number in the 
Universe} (O. Sawada and A.~Sugamoto, eds), (Japan), KEK 1979; 
R.N.~Mohapatra and G.~Senjanovic, Phys.\ Rev.\ Lett. {\bf 44}, 912 (1980)\,.

\bibitem{di_bound} S.~Davidson and A.~Ibarra,
Phy\ . Lett.\ B{\bf 535}, 25 (2002); W.~Buchmuller, P.~Di Bari and
M.~Plumacher, Nucl.\ Phys.\ B{\bf 643}, 367 (2002)\,.

\bibitem{tripletseesaw} J. Schechter and J.W.F. Valle, Phys. Rev. {\bf D 22},
2227 (1980); M.~Magg and C.~Wetterich, Phys.~Lett.~B {\bf 94},~61 (1980);
R.~N.~Mohapatra and G.~Senjanovic, Phys. Rev. D{\bf 23}, 165 (1981);
G.~Lazarides, Q.~Shafi and C.~Wetterich, Nucl. Phys.~B{\bf 181}, 287 (1981)\,.

\bibitem{ma&sarkar_prl} E.~Ma and U.~Sarkar, Phys.\ Rev.\ Lett.
{\bf 80} (1998) 5716-5719\,.

\bibitem{ma_raidal_sarkar} E.~Ma, M.~Raidal and U.~Sarkar, 
Phys.\ Rev.\ Lett.\  {\bf 85}, 3769 (2000)\,.

\bibitem{ma&sarkar_plb} E.~Ma and U.~Sarkar, 
Phys.\ Lett.\ B {\bf 638}, 356 (2006)\,.

\bibitem{sahu&sarkar_prd} S.~Antusch and S.~F.~King,
Phys.\ Lett.\ B {\bf 597}, 199 (2004); T.~Hambye and G.~Senjanovic, 
Phys.\ Lett.\ B{\bf 582}, 73 (2004); N.~Sahu and U.~Sarkar, 
Phys.\ Rev.\ D {\bf 74}, 093002 (2006); N.~Sahu and S.~Uma Sankar, 
Nucl.\ Phys.\ B {\bf 724}, 329 (2005); Phys.\ Rev.\ D {\bf 71}, 
013006 (2005)\,. 

\bibitem{singlet_resonant} M. Flanz, E.A. Paschos and U. Sarkar,
Phys.\ Lett. {\bf B 345}, 248 (1995); A.~Pilaftsis and
T.~E.~J.~Underwood, Nucl.\ Phys.\ B {\bf 692}, 303 (2004)\,.

\bibitem{triplet_resonant} G.~D'Ambrosio, T.~Hambye, A.~Hektor, 
M.~Raidal and A.~Rossi, Phys.\ Lett.\ B {\bf 604}, 199 (2004); 
E.~J.~Chun and S.~Scopel, Phys.\ Lett.\ B {\bf 636}, 278 (2006); 
[arXiv:hep-ph/0609259]\,. 

\bibitem{dark_energy} P.Q. Hung, [arXiv: hep-ph/0010126];
R.~Fardon, A.~E.~Nelson and N.~Weiner, JCAP {\bf 0410}, 005 (2004);
P.~Gu, X.~Wang and X.~Zhang, Phys.\ Rev.\ D {\bf 68}, 087301 (2003)\,.

\bibitem{cliffetal} J.~Matias and C.~P.~Burgess, JHEP {\bf 0509}, 
052 (2005)\,.

\bibitem{darkmatter}  R.~Barbieri, L.~J.~Hall and V.~S.~Rychkov,
Phys.\ Rev.\ D {\bf 74}, 015007 (2006); E.~Ma, Phys.\ Rev.\ D {\bf 73},
077301 (2006); Mod.\ Phys.\ Lett.\ A {\bf 21}, 1777 (2006);
L.~M.~Krauss, S.~Nasri and M.~Trodden, Phys.\ Rev.\ D {\bf 67}, 085002 (2003);
J.~Kubo and D.~Suematsu, Phys.\ Lett.\ B {\bf 643}, 336 (2006); 
K.~Cheung and O.~Seto, Phys.\ Rev.\ D {\bf 69}, 113009, 2004\,.

\bibitem{pascosetal.06} C.~T.~Hill, I.~Mocioiu, E.~A.~Paschos and U.~Sarkar, 
[arXiv:hep-ph/0611284]\,.

\bibitem{fukugita.86} M.~Fukugita and T.~Yanagida, Phys. Lett.
{\bf B174}, 45 (1986)\,.

\bibitem{frigerio_ma} M.~Frigerio, T.~Hambye and E.~Ma, 
JCAP {\bf 0609}, 009 (2006)\,.

\bibitem{cliff_npb.01} C.~P.~Burgess, M.~Pospelov and T.~ter Veldhuis,
  Nucl.\ Phys.\ B {\bf 619}, 709 (2001); J.~McDonald,
  Phys.\ Rev.\ D {\bf 50}, 3637 (1994)\,.

\bibitem{sahu&yajnik_plb.06} N.~Sahu and U.~A.~Yajnik, 
Phys.\ Lett.\ B {\bf 635}, 11 (2006); See, e.g., R.N.~Mohapatra and
P.B.~Pal, {\it Massive Neutrinos in Physics and Astrophysics}
(2nd edition) (World Scientific, Singapore, 1998)\,. 

\bibitem{pdg} W.M. Yao {\it et. al.} (Particle physics data group), 
Journal of Physics G 33, 1 (2006)\,. 

\bibitem{quintessence} C. Wetterich, Nucl.\ Phys.\ B {\bf 302}, 668 (1988);
P.J.E. Peebles and B. Ratra, Astrophys.\ J.\ {325}, L17 (1988)\,.

\bibitem{collider_signature} G.~Barenboim, K.~Huitu, J.~Maalampi and M.~Raidal,
  Phys.\ Lett.\ B {\bf 394}, 132 (1997); K.~Huitu, J.~Maalampi, A.~Pietila 
and M.~Raidal, Nucl.\ Phys.\ B {\bf 487}, 27 (1997); T.~Han, H.~E.~Logan, 
B.~Mukhopadhyaya and R.~Srikanth, Phys.\ Rev.\ D {\bf 72}, 053007 (2005); 
E.~Ma, M.~Raidal and U.~Sarkar, Nucl.\ Phys.\ B {\bf 615}, 313 (2001); 
C.~Yue and S.~Zhao, [arXiv: hep-ph/0701017]\,.

\bibitem{chun_plb} E.~J.~Chun, K.~Y.~Lee and S.~C.~Park, 
Phys.\ Lett.\ B {\bf 566}, 142 (2003); A.~G.~Akeroyd and M.~Aoki, 
Phys.\ Rev.\ D {\bf 72}, 035011 (2005)\,.

\end{thebibliography}
\end{document}